\documentclass[useAMS,usenatbib]{mn2e}
\usepackage[]{natbib,amsmath,amssymb,times}
\bibpunct{(}{)}{;}{a}{}{,}

\usepackage[utf8]{inputenc}
\usepackage{eurosym,graphicx,hyperref,epsfig,latexsym}

\renewcommand{\d}{\mathrm{d}}

\newcommand{\ii}{\mathrm{i}}
\newcommand{\bea}{\begin{eqnarray}}
\newcommand{\eea}{\end{eqnarray}}
\newcommand{\be}{\begin{equation}}
\newcommand{\ee}{\end{equation}}
\newcommand{\rund}[1]{\left(#1\right)}
\newcommand{\vc}[1]{\mbox{\boldmath $#1$}}

\newcommand{\eck}[1]{\left[ #1 \right]}
\newcommand{\msun}{\,h^{-1}\,M_{\odot}}

\def\elabel#1{\label{eq:#1}}

\title[Gravitational lensing effects on sub-millimetre galaxy counts]
{Gravitational lensing effects on sub-millimetre galaxy counts}
\author[Er, Li, Mao \& Cao]%
{
Xinzhong Er$^1$\thanks{E-mail: xer@nao.cas.cn},
Guoliang Li$^2$, Shude Mao$^{1,3}$, Liang Cao$^1$
\\
$^1$National Astronomical Observatories, Chinese Academy of Sciences,
20A Datun Road, Beijing 100012, China\\
$^2$Purple Mountain Observatory, Chinese Academy of Sciences,
2 West Beijing Road, Nanjing 210008, China\\
$^3$Jodrell Bank Centre for Astrophysics, University of Manchester,
Alan Turing Building, Manchester M13 9PL, UK
}%

\date{Accepted 2012 Dec 31; received 2012 Dec 20; in original form 2012 Oct 29}
\pubyear{2012}

\begin{document}

\maketitle
\begin{abstract}
We study the effects on the number counts of sub-millimetre galaxies
due to gravitational lensing. We explore the effects on the
magnification cross section due to halo density profiles, ellipticity
and cosmological parameter (the power-spectrum normalisation
$\sigma_8$).  We show that the ellipticity does not strongly affect
the magnification cross section in gravitational lensing while the
halo radial profiles do. Since the baryonic cooling effect is stronger
in galaxies than clusters, galactic haloes are more concentrated. In
light of this, a new scenario of two halo population model is
explored where galaxies are modeled as a singular isothermal sphere
profile and clusters as a Navarro, Frenk and White (NFW) profile.  We
find the transition mass between the two has modest effects on the
lensing probability. The cosmological parameter $\sigma_8$ alters the
abundance of haloes and therefore affects our results. Compared with
other methods, our model is simpler and more realistic. The
conclusions of previous works is confirm that gravitational lensing is
a natural explanation for the number count excess at the bright end.
\end{abstract}

\section{Introduction}

Sub-millimetre galaxies (SMGs) at high redshift appear to be the
counterparts of the most luminous star-forming galaxies in the local
Universe. The high luminosity of these galaxies is presumed to be the
result of large amounts of star formation, $100-1000 \msun\, {\rm
  yr}^{-1}$ and warm dust \citep{2002PhR...369..111B}.  The millimetre
and sub-millimetre wavelengths surveys have provided an important
complement to the optical and radio searches for distant galaxies. In
recent surveys, a significant population of high luminosity, high
redshift galaxies have been discovered
\citep{2006MNRAS.372.1621C,2007MNRAS.377.1557N}.  The redshift
distribution of the SMGs has a narrow distribution with a probable
median redshift of $2-3$
\citep{2005ApJ...622..772C,2007MNRAS.379.1571A}.  Surveys at
sub-millimetre wavelengths show that the SMG population has a sharp
falloff at the bright luminosity end of their luminosity
function. Gravitational lensing by intervening galaxy clusters and
groups modifies the observed number counts significantly
\citep{1996MNRAS.283.1340B, 2010ApJ...717L..31L, 2011MNRAS.411.2113J,
  2011ApJ...734...52H}. Recently it has become possible to identify the
lensed SMGs, e.g. from the ground with the South Pole Telescope (SPT)
\citep{2010ApJ...719..763V}, and from space with Herschel
\citep{2012ApJ...749...65G}.

Gravitational lensing probes the cosmology, the mass distribution in
the universe, and provides a way to study the high redshift objects
(see \citealt{2010ARA&A..48...87T} for a review). In the statistical
study of cosmological gravitational lensing, several aspects have been
studied: the lensing probability of separation of multiple images
\citep[e.g.][]{2001ApJ...549L..25K,2007MNRAS.378..469L}; the number of
giant arcs formed by gravitational lensing
\citep[e.g.][]{1998A&A...330....1B,2005ApJ...635..795L,2011MNRAS.418...54H}
and the modified luminosity function (LF) of background sources
\citep[e.g.][]{2010ApJ...717L..31L,2011ApJ...734...52H,
  2011Natur.469..181W,2012arXiv1205.3778W}.

In this paper, we will focus on the LF of high redshift SMGs. There
are a relatively large number of SMGs, and their redshift is likely
high ($z>2$), both of which help to create an excellent source
population for lensing studies \citep{2005ApJ...622..772C}. To predict
the observed source counts, the intrinsic LF needs to be known.
Furthermore, in order to determine the probability (total cross
section) of gravitational lensing, we need to know several properties
of lens haloes, e.g. their abundance and internal structure, which are
affected by the cosmological parameters of the universe. More
specifically, the cross section due to gravitational lensing may be
affected by halo ellipticity
\citep[e.g.][]{2001ApJ...553..709R,2005ApJ...624...34H}, the radial
profile of the lens halo
\citep{2002ApJ...566..652L,2004ApJ...610..663O} as well as the size of
background galaxies \citep{2002MNRAS.329..445P,
  2011ApJ...734...52H}. In \citet{2010MNRAS.406.2352L}, haloes are
modelled with NFW profiles, while ellipticity is added to their
lensing potential to increase the lensing probability.  More recently,
\citet{2012ApJ...755...46L}, considered a composite model for galactic
sized halo, where dark matter is modelled as NFW and the stellar
component by a Sersic profile \citep{1963BAAA....6...41S}. It is in
fact close to the isothermal profile. They also use the public code
GLAFIC from \citet{2010PASJ...62.1017O} to study the effect of halo
ellipticity on the lensing cross section and find that the ellipticity
only weakly affects the cross section of the isothermal halo lens.
Our analytical results confirm their finding.

Axisymmetric lensing models offer simplicity in the study of lens
statistics. Several spherical lenses have been studied, such as the
Singular Isothermal Sphere (SIS) and the Navarro-Frenk-White profile
\citep[e.g.][]{2002ApJ...566..652L,2004ApJ...610..663O}. Due to
baryonic cooling, clusters and galaxies will have different mass
profiles. Therefore, a combination of two population halo mass
profiles will be studied in this work.  The statistical effect of
lensing will be affected by the cooling mass scale $M_{\rm cool}$
\citep[see, e.g., the study on multiple images and image splitting
  by][]{2002ApJ...566..652L,2004A&A...418..387C}. A cooling mass scale
of $\sim 10^{13} \msun$ has been suggested \citep[see, e.g.,
][]{2000ApJ...532..679P,2001ApJ...559..531K}. On large scales the mass
function of dark haloes relates to the primordial fluctuations of the
universe, which can be characterised by the power-spectrum
normalisation parameter $\sigma_8$. Thus, the total cross section will
be sensitive to this parameter as well.  Moreover, previous studies
have shown that different aspects also affect the lensing efficiency,
like the substructures within the dark haloes
\citep{2006MNRAS.367.1241O}, central massive black holes
\citep{2012MNRAS.420..792M,2012MNRAS.419.2424L} and external shear
\citep{2005ApJ...624...34H}. Most of them probably do not play
significant roles, and so they will not be considered in this paper.

We will study the lensing efficiency, i.e. the magnification cross
section in this paper. We start from a single lens halo for different
halo profiles, and present the lensing cross section dependence on
halo properties. We show that the halo ellipticity does not affect the
cross section significantly, while the halo density profile does.
As we mentioned before, a combination of halo profiles is more
physical and we find such a scenario can produce a higher lensing
probability than a single universal NFW profile. We present our
calculation and employ it to study the luminosity function in Section
3; we further discuss our results in Section 4.  The cosmology that we
adopt in this paper is a $\Lambda$CDM model with parameters based on
the results of the Wilkinson Microwave Anisotropy Probe seven year
data \citep{2011ApJS..192...18K}: $\Omega_{\Lambda}=0.734$,
$\Omega_{\rm m}=0.266$, $\Omega_{\rm b}=0.0449$, $n=0.963$, a Hubble
constant $H_0 = 100 h$ km\,s$^{-1}$\,Mpc$^{-1}$ and $h=0.71$. We allow
the $\sigma_8$ parameter to vary, but use the WMAP7 value
$\sigma_8=0.8$ if not mentioned.

\section{Basic Formalism}

The fundamentals of gravitational lensing can be found in
\citet{2001PhR...340..291B}. For its elegance and brevity, we shall
use the complex notation. The thin-lens approximation is adopted,
implying that the lensing mass distribution can be projected onto the
lens plane perpendicular to the line-of-sight. We introduce angular
coordinates $\vc\theta$ with respect to the line-of-sight. The lensing
convergence, that is the dimensionless projected surface-mass density,
can be written as
\be
\kappa(\vc\theta) = \Sigma(\vc\theta)/\Sigma_{\rm cr},\;\;\; {\rm where} \;\;\;
\Sigma_{\rm cr} = \frac{c^2}{4\pi G} \frac{D_{\rm s}}{D_{\rm d} D_{\rm ds}}\;
\ee
is the critical surface mass density depending on the angular-diameter
distances $D_{\rm s}$, $D_{\rm d}$ and $D_{\rm ds}$ from the observer
to the source, the observer to the lens, and the lens to the source,
respectively.  $\Sigma(\vc\theta)$ is the projected surface-mass
density of the lens.  All lensing quantities can be derived from the
effective lensing potential $\psi$,
\be
  \psi(\vc\theta) = \frac{1}{\pi}\int_{{\cal R}^2} \d^2\theta'\kappa(\vc\theta')\;
  {\rm ln}|\vc\theta-\vc\theta'|\;.
\ee
To the lowest order, image distortions caused by gravitational lensing are
described by the complex shear
\be
\gamma = \frac{1}{2}\left(\partial_1^2\psi-\partial_2^2\psi\right)
+ {\rm i}\partial_1\partial_2\psi\;.
\label{shear}
\ee
The magnification for a point source is given by
\be
\mu = \dfrac{1}{(1-\kappa)^2 \,-\,|\gamma|^2}.
\elabel{anamu}
\ee

\subsection{Lensing properties of different dark matter halo profiles}

Having laid out a general formalism, we will present the
basic lensing properties of two different dark matter halo profiles
that will be used in this paper (see below). They are
Singular Isothermal Sphere (SIS) and Navarro-Frenk-White (NFW,
\citealt{1997ApJ...490..493N}) profiles.

The dimensionless surface mass density and shear for an SIS halo are
\be
\kappa = \frac{\theta_{\rm E}}{2 \theta}\;,\;\;\;\;\;
\gamma = -{\theta_{\rm E} \over 2 \theta}{\rm e}^{2\ii \phi},
\elabel{kappasis}
\ee
where $\phi$ is the position angle around the lens, $\theta
=\sqrt{\theta_1^2 + \theta_2^2}$ is the angular separation, and
$\theta_{\rm E}$ is the Einstein angular radius, which is calculated
by
\be
\theta_{\rm E} = 4 \pi \rund{\sigma_v \over c}^2 {D_{\rm ds} \over D_{\rm s}},
\elabel{thetaE}
\ee
where $\sigma_v$ is the one-dimensional velocity dispersion, and $c$
is the speed of light. In the rest of the paper, $c$ is used as the
concentration parameter. The lensing properties of the NFW profile
has been calculated by \citet{1996A&A...313..697B}. The convergence
is analytically given by
\be
\kappa = 2\kappa_s{f(x)\over x^2-1},
\label{kappanfw}
\ee
where $x=\theta D_{\rm d}/r_s$ (or $x=\theta/\theta_s, \theta_s=r_s/D_{\rm d}$) is the
dimensionless radius, and the function $f(x)$ is defined as
\be
f(x)=
\begin{cases}
  1 - \dfrac{{\rm arcsech} x }{\sqrt{1 - x^2}} \;\; (x<1); \\
  \\0 \quad\quad\quad\quad\quad \;\;\; (x=1); \\ \\
  1 - \dfrac{{\rm arcsec} x }{\sqrt{x^2 - 1}} \;\; (x>1).\\
\end{cases}
\label{eq:fx}
\ee
The physical properties of the halo are contained in the parameter
$\kappa_s=\rho_{\rm crit} \Delta_c r_s/\Sigma_{\rm cr}$, where
$\Delta_c$ is the dimensionless characteristic density and $\rho_{\rm
  crit}$ is the critical density. The halo mass is defined as
$M_{200}=800/3\rho_{\rm crit}\pi r_{200}^3$, where $r_{200}=r_s\,c$
and $c$ is the concentration parameter (see the appendix in Navarro
et~al. 1997).  More lensing properties of the NFW halo profile can be
found in \citet{2000ApJ...534...34W}.

\subsection{Lensing cross sections of dark matter haloes}

\begin{figure}
\centerline{\scalebox{1.0}
  {\includegraphics[width=9.5cm,height=8.0cm]{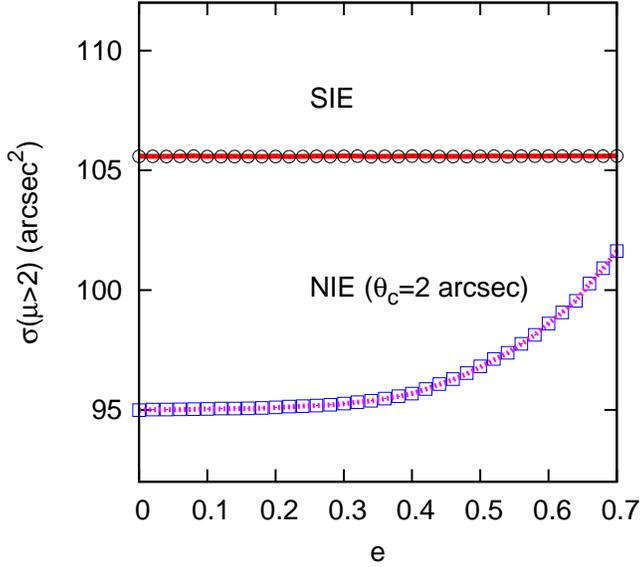}}}
\caption{Cross section $\sigma(\mu>2)$ vs. halo ellipticity for the
  SIE and NIE halo profiles. The solid line is the analytical result
  for the SIE halo given by Eq.~\ref{eq:crosssie}. The open circles
  and dashed line are the cross section using cell counts in the
  magnification map, where the magnification is derived by
  Eqs.~\ref{eq:siemag} and \ref{eq:emag} for SIE and NIE model
  respectively. The open squares represent the numerical results for
  the NIE halo which magnification is derived numerically (Appendix
  B).  $\theta_{\rm E}=5.5$ arcsec is used here for all cases (which
  corresponds to the halo mass of $M_{200}=10^{14} \msun$). For the
  NIE halo, a core radius $\theta_c=2$ arcsec is used.}
\label{fig:effellp}
\end{figure}
\begin{figure}
\centerline{\scalebox{1.0}
  {\includegraphics[width=9cm,height=8.0cm]{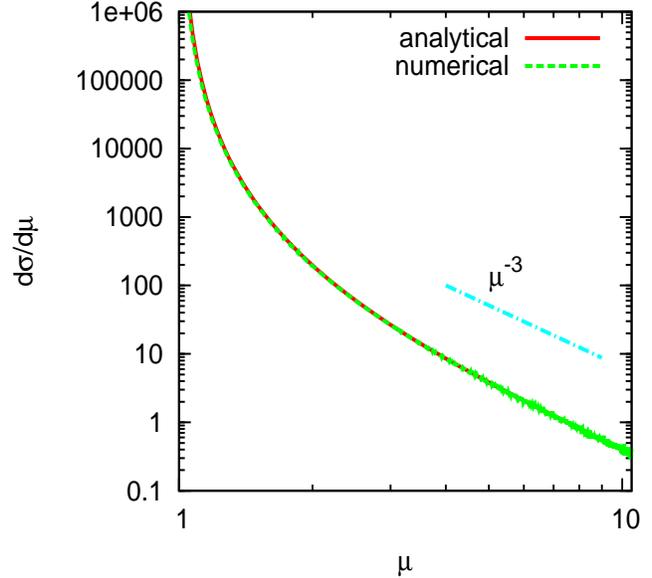}}}
\caption{Differential lensing cross section as a function of
  magnification $\mu$ for the SIE halo. $\theta_{\rm E}=5.5$ arcsec is
  used as in Fig.~\ref{fig:effellp}. The solid and dashed lines are
  the analytical and numerical results, which almost overlap. The
  dot-dashed line shows the expected asymptotic behaviour.}
\label{fig:effmu}
\end{figure}

We will calculate the cross section of different halo profiles, and
present their dependence on different parameters, e.g. halo
ellipticity and concentration etc. Some previous studies have shown
that the magnification cross section increases with halo ellipticity
dramatically \citep[e.g.][]{2010MNRAS.406.2352L}. We, however, find that
this is mainly because the halo mass is also changed.  We start with
an elliptical surface density for which the mass within an ellipse is
identical with that of the axis-symmetric one. In general, the
corresponding magnification can be calculated numerically
\citep{1990A&A...231...19S}.


We first perform a series of tests for the cross section of the halo
ellipticity. The halo ellipticity is characterised by $e=1-q$, where
$q=\theta_a/\theta_b$ is the axis ratio, $\theta_a$ and $\theta_b$ are
the major and minor axes, respectively. The angular separation
$\theta$ will be replaced by $\Theta=\sqrt{q\theta_1^2+\theta_2^2 /q}$
(uppercase $\Theta$ is used for the elliptical coordinate) in
Eq.\,\ref{eq:kappasis} to calculate $\kappa$ and $\gamma$ for a
Singular Isothermal Ellipsoid (SIE) halo (see
\citealt{1998ApJ...495..157K} for more details on the lensing
properties of an SIE halo). It is a radial coordinate and will be
constant on elliptical contours (thus $x=\Theta D_{\rm d} /r_s$ in the
NFW model). The total mass within a given $\Theta$ is invariant with
the ellipticity $e$. Following Eq.~\ref{eq:emag}, the magnification of
an SIE halo has an analytical expression
\be
\mu = {\Theta \over \Theta - \theta_{\rm E}}.
\elabel{siemag}
\ee
For a given magnification $\mu_{\rm min}$, the cross section $\sigma(\mu)$
is the area inside which the magnification of a source is equal to or
larger than $\mu_{\rm min}$ on the source plane. For the SIE halo,
the cross section can be given analytically
\be
\sigma (\mu)= \pi \theta_{\rm E}^2 \rund{{1\over (\mu-1)^2} +
{1\over (\mu+1)^2}}.
\elabel{crosssie}
\ee
%
Notice that at large magnification, the cross section follows the
predicted asymptotic power-law $1/\mu^2$ for $\mu \gg 1$
\citep{1992grle.book.....S}.

The mass of the SIE halo is related to $\theta_{\rm E}$.  For the SIE
model the mass is related to the rotation velocity, which can be
calculated using $M_{200}=V_c^3/(10GH(z))$
\citep{1998MNRAS.295..319M}. The relation between the rotation
velocity and velocity dispersion is complicated and may be
different for different types of galaxies. We use the approximate relation
between the rotation velocity and the velocity dispersion
$\sigma_v=V_c/\sqrt{2}$, which is suggested by
\citet{2010MNRAS.402.2031C}.
The cross section has an expression of mass and distance
\bea
\sigma (\mu) &=& {16 \pi^3 \over c^4} \rund{25\over 2}^{2/3}
\rund{G H M_{200}}^{4/3}\rund{D_{\rm ds} \over D_{\rm s}}^2\nonumber\\
&&\rund{{1\over (\mu-1)^2} + {1\over (\mu+1)^2}}.
\eea
In our definition of the ellipticity, the mass and the cross section
do not change with ellipticity. We can see that only the halo mass and
the redshifts of lens and source affect the cross section. In the
appendix, one can find more analytical results for the
magnification and cross section of SIE and Non-singular Isothermal
Ellipsoid (NIE) profiles.

To check the accuracy of Eq.~\ref{eq:crosssie}, we create a magnification
map using Eq.~\ref{eq:siemag} and sum up the pixels following
Eq.~\ref{eq:ecross}.  A halo with mass of $M_{200}=10^{14}\msun$ is
used, which gives $\theta_{\rm E}=5.5$ arcsec for $z_{\rm d}=0.5$ and
$z_{\rm s}=2.0$. The resolution of the magnification grid is
$\Theta_{\rm E}/500$. We increase the lens halo ellipticity to see its
effect to the cross section. For simplicity, we use a constant halo
ellipticity as a function of radius.  The numerical results are shown
by the open circles in Fig.~\ref{fig:effellp} and the red solid line
shows the prediction of Eq.~\ref{eq:crosssie}. We can see that the
results agree well with each other. It encourages us to explore whether
such an independence is also valid for other kinds of mass
distribution models. Fig.~\ref{fig:effmu} presents the probability of
cross section as a function of magnification for the SIE halo ($\d
\sigma/\d \mu$). The solid line represents the theoretical prediction,
while the dashed line is obtained by numerical calculation. One can see
that the numerical result closely follows the theoretical
prediction. As expected, at small $\mu$ the probability density
distribution rapidly decreases; at large $\mu$, it becomes close to
the asymptotic relation $\propto \mu^{-3}$.

By adding a core of $\theta_c=2$ arcsec into the SIE model,
we extend our test to the NIE model. In Appendix A, we show an analytical
expression of the magnification (Eq.~\ref{eq:emag}).
By summing up the pixels in the magnification map as before, we show
the ``theoretical'' cross section as a dashed line in
Fig.~\ref{fig:effellp}. We call it the theoretical cross section
because the magnification is calculated analytically. We also show the
result where the magnification is calculated numerically. One can
see that the cross section of NIE halo increases with the ellipticity
by a few percent for the parameters considered here.

For most elliptical mass distribution models, the magnification
can not be derived analytically, e.g., the Einasto model
\citep{2012A&A...546A..32R}, the Hernquist model
\citep{2002A&A...393..485B} and the NFW
model. \citet{1990A&A...231...19S} proposed a way to calculate the
lensing properties for any kind of elliptical mass model. This
algorithm is summarised in \citet{2001astro.ph..2341K} and
revised for our definition of elliptical coordinate in Appendix
B. Following this approach, we recalculate the magnification numerically
for the NIE halo and plot the results in Fig.~\ref{fig:effellp} as the open
squares. The good agreement between the dashed line and the open
squares guarantees the accuracy of the numerically calculated
magnification and confirms the weak dependence between the cross
section and the ellipticity.


\begin{figure}
  \centerline{
  \includegraphics[width=9.5cm,height=8cm]{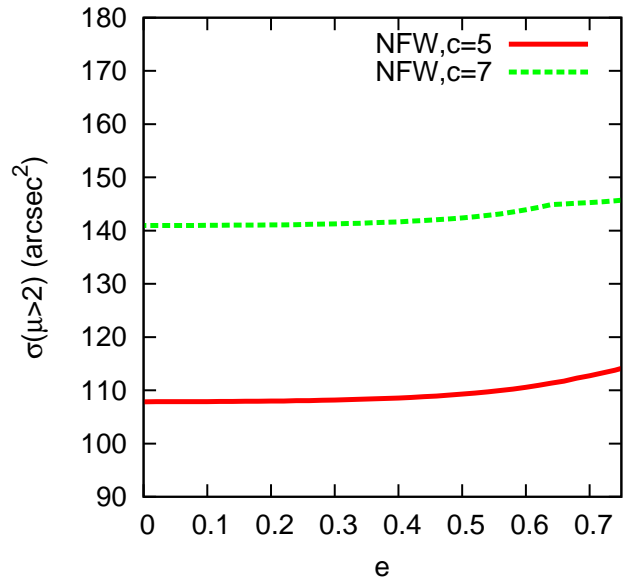}}
\caption{Cumulative cross section $\sigma(\mu>2)$ vs. halo ellipticity
  for the NFW halo profiles: The solid (dashed) line is the result for
  the NFW halo profile with concentration $c=5$ ($c=7$).  The same
  condition, halo mass $M_{200}=10^{14}\msun$ and redshift $z_d=0.5$,
  $z_s=2.0$ are used for all the figures in this section.}
\label{fig:nfwellp}
\end{figure}
\begin{figure}
\centerline{\scalebox{1.0}
  {\includegraphics[width=9.5cm,height=8cm]{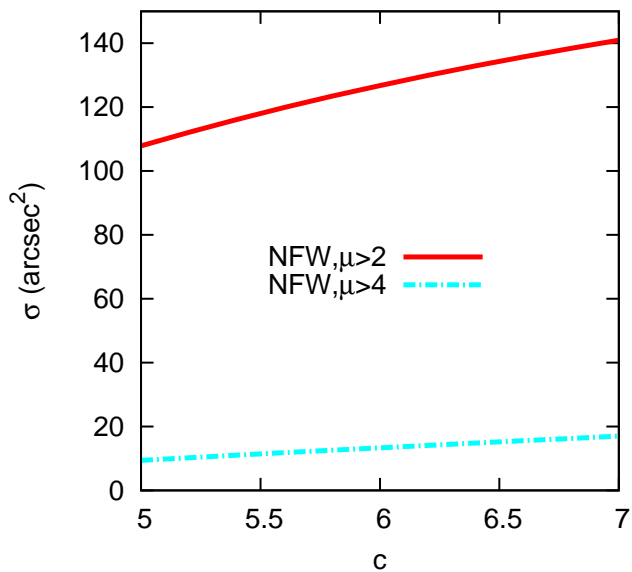}}}
\caption{Lensing cross section vs. concentration for the NFW
  profiles. The solid line is the normalised cross section of $\mu>2$
  for NFW profiles. The dotted line is for the cross section with
  $\mu>4$. The same halo mass $M_{200}=10^{14}\msun$ is used.}
\label{fig:nfwc}
\end{figure}
\begin{figure}
\centerline{\scalebox{1.0}
  {\includegraphics[width=9cm,height=8cm]{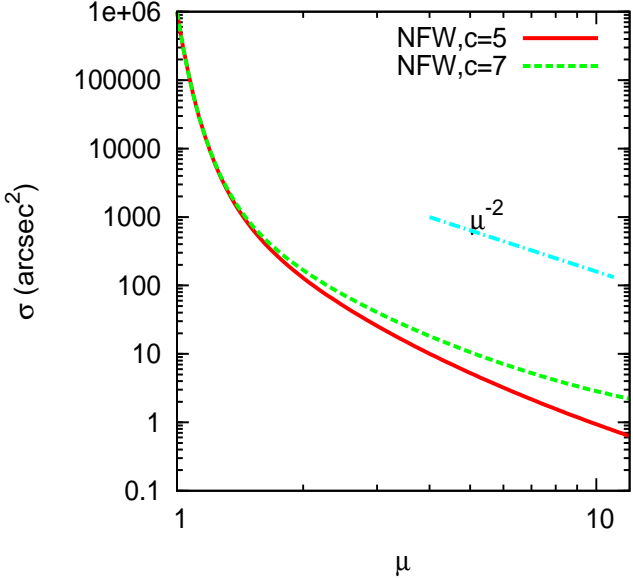}}}
\caption{Lensing cross section vs. $\mu$ for the NFW profile.
The solid and dashed lines represent the results for the NFW profile with
$c=5,7$. Again a halo mass $M_{200}=10^{14}\msun$ is used.}
\label{fig:nfwmu}
\end{figure}

For an elliptical NFW (eNFW) halo, the lensing properties,
e.g. shear, magnification and the cross section, can be calculated
numerically as above.
We perform similar numerical tests for the cross section $\sigma(\mu)$
using eNFW halo profiles. The same mass $M_{200}=10^{14}\msun$ and
redshifts ($z_d=0.5$, $z_s=2.0$) are used.

In Fig.~\ref{fig:nfwellp}, we show the cross section $\sigma(\mu>2)$
variation with halo ellipticity. Two different concentration
parameters $c=5,7$ are used for NFW halo profiles. We find that there
is a weak dependence on the halo ellipticity of $\sigma$ (less than
$5$ percent).
Fig.~\ref{fig:nfwc} shows the cross section variation with the concentration
parameter $c$ for the NFW profiles. As expected, with a higher
concentration the lensing efficiency is higher.
Fig.~\ref{fig:nfwmu} shows the cross section
$\sigma(\mu)$ dependence on the magnification $\mu$. At high
magnification, there is a significant difference in cross section between two
concentration parameters for the NFW profile.

There are a number of complications we have ignored (see also the
discussion). The numerical tests we performed only take into account
the projected shape, and assumed the ellipticity is constant as a
function of radius. In reality, this may not be true. In particular,
irregular shape lens haloes may have different cross sections and
probability of generating multiple images especially for the massive
haloes not long after the merging event. Other properties, such as the
substructures may also affect the cross section.  The finite source
size may smooth the magnification and lowers the lensing efficiency
(see the end of Section 3). Thus, in reality, the lensing halo may
have slightly different cross sections, which need to be further
studied using numerical simulation, although the required spatial and
mass resolutions may be challenging. As the effect of ellipticity on
the lensing cross section seems to be relatively small, in this paper,
for simplicity we will adopt the spherical lens model in the following
calculations.

\subsection{Probability function of magnification}

The probability distribution $P(\mu)$ can be estimated either by
ray-tracing simulations \citep[e.g.][]{2007MNRAS.382..121H}, or by
semi-analytical methods \citep[e.g.][]{2010ApJ...717L..31L}, which
integrates all the halo contributions along the line of sight from us
to the source redshift. We will adopt the latter approach,
which allows us to easily test the dependence on
parameters, e.g. $\sigma_8$ and $M_{\rm cool}$.

Different halo density profiles will be employed for lens haloes,
i.e. SIS and NFW. The mass-concentration relation for the NFW halo
profile has been studied by several authors
\citep[e.g.][]{2001MNRAS.321..559B, 2009ApJ...707..354Z}. In this
paper, we adopt the simple model
\be
c(M,z)={9 \over 1+z} \rund{M\over M_*}^{-0.13},
\elabel{masscrelation}
\ee
where $M_*$ is calculated by $\sigma_m(M_*) = \delta_c$.
$\sigma_m^2(M)$ is the variance of the linear density field
(Eq. \ref{eq:linearpow}), and $\delta_c$ is the linearly extrapolated
density contrast threshold at redshift $z$ in spherical collapse,
here we use $\delta_c(z=0)=1.686$.

The halo mass function \citep{1974ApJ...187..425P}, which determines
the number of haloes given a mass at each redshift $n(M,z)$, can be
written as
\be
{\d n\over \d M}\d M = {\bar\rho \over M} f(\nu) \d \nu,
\ee
where $\bar\rho$ is the comoving mean matter density of the Universe
and $\nu=\delta_c/\sigma_m(M)$. In the
\citet{1999MNRAS.308..119S} formalism, we have,
\be
\nu f(\nu)=A\sqrt{\dfrac{2}{\pi}a\nu^2}\eck{1+(a\nu^2)^{-p}}
    {\rm exp}\eck{-a\nu^2/2}.
\elabel{massfuncmu}
\ee
Here $\sigma_m^2(M)$ is the variance of the linear density field in a top hat of
radius $r$ that encloses $M=4\pi r^3 \rho_m/3$ at the background density
\be
\sigma_m^2(r) = \int \dfrac{\d^3k}{(2\pi)^3} |W(kr)|^2 P_L(k),
\elabel{linearpow}
\ee
where $P_L(k)$ is the linear power spectrum
\be P_L(k,z) \propto  k^n  D^2(z) T^2(k), \ee
and $W(kr)$ is the Fourier transform of the top hat window function. The
fitting formula of the linear transfer function including baryons
\citep{1998ApJ...496..605E} is used here and we use the WMAP 7-year cosmology
\citep{2011ApJS..192...18K}.
Alternative models for the shape of $n(M,z)$
are available in the literature \citep[e.g.][]{2000ApJ...541...10M},
we will not consider these since Eq. (\ref{eq:massfuncmu}) provides a good
description of the mass function in numerical simulations. We
however use another form of Eq. (\ref{eq:massfuncmu}) which is easier
to implement in numerical integration
\be f(\sigma,z) \equiv {M\over
\bar\rho}{\d n \over \d {\rm ln}\sigma^{-1}}.
\ee
It has a fitting formula of
\be
f(\sigma) = 0.315 \, {\exp}(-|{\ln}\,\sigma^{-1} + 0.61|^{0.38})
\ee
in the range $-1.2\leq {\ln}\,\sigma^{-1}\leq
1.05$ \citep{2001MNRAS.321..372J}.

The lensing cross section $\sigma(>\mu,z_d,M)$ for a single halo will be
calculated as a function of mass. The halo mass is used as $M_{200}$. The
cross section of the SIS halo can be calculated analytically
(Eq.~\ref{eq:crosssie}) and that of the NFW halo will be performed
numerically.  The sum of all cross sections in the Universe can be
written as \citep{2010MNRAS.406.2352L}
\be
\sigma_{\rm tot}(\mu) = 4\pi \int {D_A^2(z_d) \over H(z_d)}\,\d z_d
\int \d M {\d n(z_d,M) \over \d M}\;\sigma(>\mu,z_d,M),
\elabel{cstotal}
\ee
where $D_A$ is the comoving angular diameter distance, and we make a
simplification that the source SMGs are distributed at a fixed
redshift. We place the sub-millimetre source galaxies mostly at
redshift $z_s=3.0$ \citep{2005ApJ...622..772C,2007ApJ...658..778Y},
but will later allow it to vary between $2-4$ (see Section 3). The
lens haloes are distributed between the sources and us
($0.001<z_d<1.5$).

The total cross section is integrated using Eq.~\ref{eq:cstotal}, then
the cumulative probability that a source at $z_s$ is magnified by a
factor greater than $\mu_{\rm min}$ is then
\be
P(\mu) = {\sigma_{\rm tot} (\mu)\over 4 \pi }.
\ee
The probability density can be obtained by $p(\mu)=-\d P(\mu) /\d\mu$.
The magnification probability function will be affected by several
aspects, e.g. the lens halo profile, halo mass function, and the lens
and the source redshifts etc. We will study several factors.  First of
all, the cross section of a single halo $\sigma(>\mu,z_d,M)$ can be
calculated using different models. We perform the calculation for
three profiles: 1) the SIS halo profile, 2) the NFW halo profile, and
3) a two population combination of SIS and NFW profiles where the
transition occurs at the cooling mass scale $M_{\rm cool}$ between
galaxies and clusters.

\begin{figure*}
\centerline{\scalebox{1.0}
  {\includegraphics[width=7.0cm,height=6.4cm]{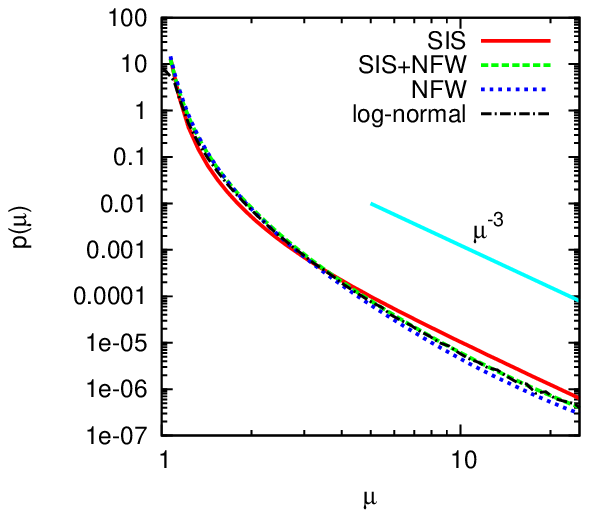}
    \includegraphics[width=5.5cm,height=5.1cm]{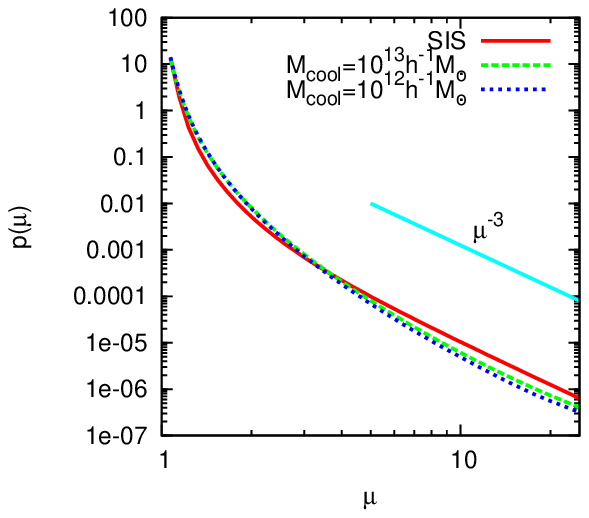}
    \includegraphics[width=5.5cm,height=5.1cm]{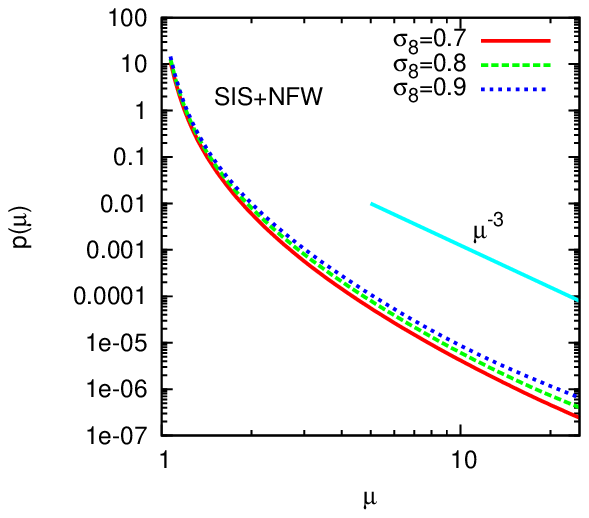}}}
\caption{Lensing probability density $p(\mu)$ along the line of sight
  up to $z_s=3.0$ from all intervening haloes. Left: Results using
  different halo profiles of SIS, SIS+NFW and NFW are shown as the red
  solid, green dashed, and blue short dashed lines
  respectively. $M_{\rm cool}=10^{13}\msun$ is used. The black
  dot-dashed line represents the combined probability of log-normal
  approximation \citep{2011A&A...536A..85H} and SIS+NFW. Middle:
  Results for the SIS+NFW halo profile using different $M_{\rm
    cool}$'s: $10^{13} \msun$ (green dashed line), $10^{12}\msun$
  (blue short dashed line). Right: Results for SIS+NFW halo profile
  using different $\sigma_8$'s: $\sigma_8=0.7$ (red solid line),
  $\sigma_8=0.8$ (green dashed line), $\sigma_8=0.9$ (blue short
  dashed line). In all three panels, the green dashed line represents
  the result using the same parameters (SIS+NFW halo, $M_{\rm
    cool}=10^{13}\msun$ and $\sigma_8=0.8$) for better comparisons.}
\label{fig:muprob}
\end{figure*}

In the left panel of Fig.~\ref{fig:muprob}, one can see that the lens
halo profile will affect the probability function $ p(\mu)$. All the
results follow $p(\mu)\propto \mu^{-3}$ as expected when $\mu \gg 1$.
At small $\mu$, all the profiles generate similar lensing probability
and all the curves drop rapidly with increasing $\mu$.
The SIS model will generate a larger probability for large $\mu$;
when $\mu$ approaches to $1$, the SIS model has a slightly smaller
probability than the NFW profile. The result using SIS+NFW however is
close to that of the NFW profile. The reason is that the main
difference between SIS and NFW is for large $\mu$, and the main
contribution to large $\mu$ is from massive haloes, i.e. the NFW
haloes here.

As expected, the transition mass $M_{\rm cool}$ also affects the
probability function. For a larger $M_{\rm cool}$, there will be more
lenses modeled as SIS, as a result one will obtain a higher
probability at large $\mu$ (middle panel in Fig.~\ref{fig:muprob}). In
particular, we find that the probability is about $10\%$ lower if we
use $M_{\rm cool}=10^{12}\msun$ at large $\mu$ than the case with
$M_{\rm cool}=10^{13}\msun$; at small $\mu$ there is a few percent
difference.
In the right panel of Fig.~\ref{fig:muprob}, one can see that a larger
$\sigma_8$ increases the lensing probability due to a larger number of
massive structures. The difference exists for most $\mu$, although it
is more significant for large $\mu$ ( $\sim25\%$) than for smaller
$\mu$ ($\sim10\%$).

\begin{figure}
\centerline{\scalebox{1.0}
{\includegraphics[width=9cm, height=8cm]{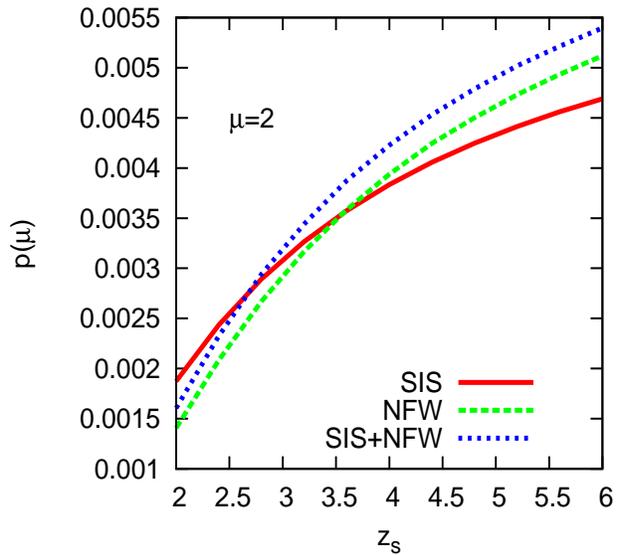}}}
\caption{Lensing probability density $ p(\mu=2)$ from intervening
  haloes ($0.001 < z_d <1.5$) for different source redshift $z_s$. The
  solid, dashed and dotted lines represent the probabilities for the SIS,
  NFW and SIS+NFW profiles respectively. }
\label{fig:muzs}
\end{figure}

In addition, we use different source redshifts. The same lens
halo redshifts are used ($0.001<z_{\rm d}<1.5$), since at high
redshift, the number density of lens halo will dramatically
decrease. In Fig.~\ref{fig:muzs}, we can see that the lensing
probability increases with the source redshift; at large redshift it
increases more slowly. At very high redshifts (e.g., $z_s>5.0$),
the luminous source number density is low, thus high redshift sources
will not strongly affect our result. Moreover, different halo
profiles have different source redshift dependence. The probability of
NFW profiles increase faster than that of SIS profile.
%

\section{Number counts of sub-millimetre galaxies}

The intrinsic number density distribution of a population of galaxies
can be fitted by empirical or semi-analytical models
\citep{2005MNRAS.356.1191B}.
In this paper we adopt a \citet{1976ApJ...203..297S} form
for the intrinsic luminosity function:
\be
{\d n \over \d S} = {n^* \over S^*} \rund{S\over S^*}^{\alpha}
{\rm e}^{-S/S^*},
\ee
where $n^*$, $S^*$, and $\alpha$ are free parameters. Lensing by
intervening haloes changes the intrinsic $\d n/\d S$ to its observed
counterpart. In addition to making sources appear brighter,
gravitational lensing also dilutes the source number density by
magnifying the observed solid angle. As discussed in
\cite{2011MNRAS.411.2113J}, the observed number counts for the whole
population is
\be
{\d n \over \d S} = {1\over \langle \mu\rangle }\int {1\over \mu' }
{\d p \over \d\mu'} {\d n' \over \d S'}\rund{S'=S/\mu'} \d\mu'.
\label{obsnumber}
\ee
The probability at small magnification is difficult to calculate using
the halo model. We perform the integral (Eq.\ref{obsnumber}) from
$\mu_{\rm min}=1.1$ to $\mu_{\rm max}=30$, and use probability
$(1-P_0)\delta(\mu=1)$ for $\mu<\mu_{\rm min}$, where $P_0$ is the
cumulative probability in image plane from $\mu_{\rm min}$ to
$\mu_{\rm max}$, and $\delta(\mu=1)$ is the Dirac function. We discuss
how the choice of $\mu_{\rm min}$ affects the results at the end of
this section.

\begin{figure}
\centerline{\scalebox{1.0}
  {\includegraphics[width=9.5cm,height=8cm]{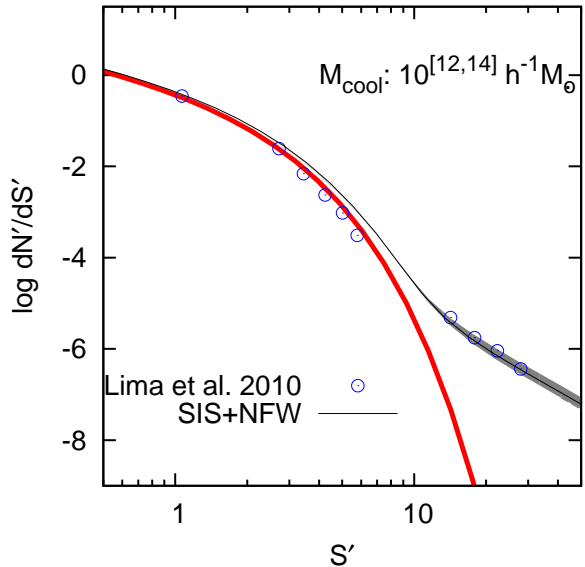}}}
\caption{The scaled intrinsic and lensed galaxy number counts $\d
  N'/\d S'$: the solid red line is the intrinsic Schechter luminosity
  function. The black line with grey shaded region represents the lensed
  galaxy number count due to the combination of halo profiles for
  sources at redshift $z_s=3.0$, and the shaded region displays the
  range of lensing predictions due to different transition masses
  $M_{\rm cool}$. The open circles are data points from the right panel of
  Fig.\,2 in \citet{2010ApJ...717L..31L}. }
\label{fig:lima}
\end{figure}

First, we compare our prediction with the results given in the right
panel of Fig.\,2 in \citet{2010ApJ...717L..31L}. We used the rescaled
intrinsic Schechter function
\be
{\d N'\over \d S'} = S'^{\alpha} {\rm e}^{-S'},
\ee
where $N'=n/n^*$ and $S'=S/S^*$. Different $n^*$ and $S^*$ are used
for different wavelengths \citep[for more detail
  see][]{2010ApJ...717L..31L}. We adopt a combination of spherical NFW
and SIS halo profiles for lens and show our prediction in
Fig.~\ref{fig:lima}. The rescaled luminosity functions before and
after lensing are shown by the lines and the points are the rescaled
data using different wavelengths and different surveys.  One can see
that our result can match the data as well as the model prediction of
\citet{2010ApJ...717L..31L}. From the view of methodology, our model
is simpler and more meaningful. Their model has to adopt a
high ellipticity of halo (e.g. $0.4$) to reach the required lensing
efficiency. We also show the effect of the transition mass $M_{\rm
  cool}$. We allow $M_{\rm cool}$ to vary from $10^{12}\msun$ to
$10^{14}\msun$. The predicted uncertainty is shown by the shaded
region. The model with a larger $M_{\rm cool}$ will predict a higher
number count at the bright end due to more effective SIS halo lenses.

\begin{figure}
  \centerline{\scalebox{1.0}
    {\includegraphics[width=13cm,height=9cm]{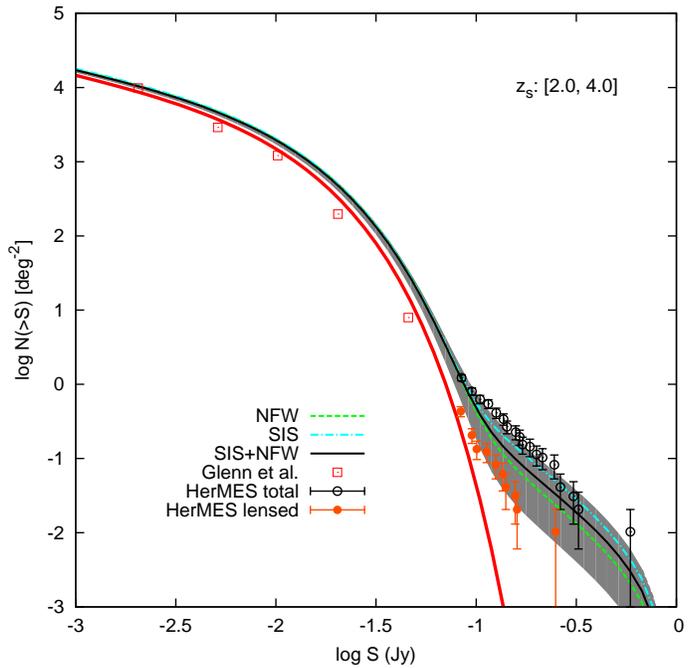}}}
  \caption{Intrinsic and lensed galaxy number counts: the red thick
    solid line is the intrinsic Schechter luminosity function
    describing galaxy number counts. The other lines represent the
    lensed galaxy number counts due to different lens halo profiles:
    SIS (cerulean dot-dashed line), NFW (green dashed line) and
    SIS+NFW (dark solid line), assuming that all the source galaxies
    are at redshift $z_s=3.0$. The shaded region displays the range of
    prediction due to different source redshifts (from $z_s=2.0$ to
    $z_s=4.0$).  The points are cumulative $500$ $\mu$m number counts
    for HerMES blank-field catalogs
    \citep{2010A&A...518L..21O,2012arXiv1203.2562H} and $P(D)$
    analysis \citep{2010MNRAS.409..109G}.}
  \label{fig:lmbig}
\end{figure}

\begin{figure}
  \includegraphics[width=9.5cm,height=8cm]{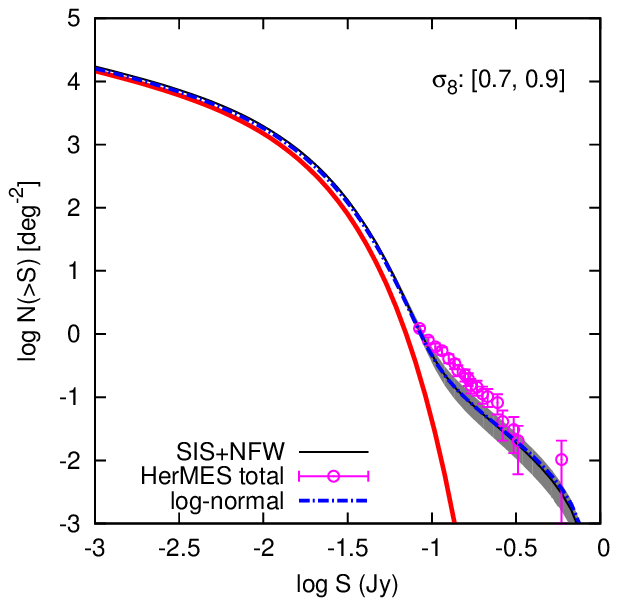}
\caption{Similar to Fig.~\ref{fig:lmbig}, comparison of different
  $\sigma_8$'s: the solid dark line is the result of SIS+NFW halo with
  $\sigma_8=0.8$ assuming $z_s=3.0$ and $M_{\rm
    cool}=10^{13}\msun$. The shaded region displays the range of
  predictions due to different $\sigma_8$ values (from $\sigma_8=0.7$
  to $\sigma_8=0.9$). The blue line represents the result of
  combination of log-normal approximation and SIS+NFW (see end of
  section 3 for more detail).}
\label{fig:lm}
\end{figure}

We then apply our method to calculate the sub-millimetre galaxy counts
of HerMES observation.  We use $n^*= 5\times 10^3 /$deg$^2$, $S^*=10$
mJy and $\alpha=-1.0$ obtained by fitting the low flux data points
\citep{2010MNRAS.409..109G}. In Figs.~\ref{fig:lmbig} and \ref{fig:lm}
the lines show the luminosity function before and after lensing. Here
we mainly consider a source redshift $z_s=3.0$. The SIS profile will
generate more lensed images than the NFW profile. The points show the
HerMES data \citep{2012arXiv1203.2562H} to compare with our
predictions. Similar as Fig.~\ref{fig:muprob}, different models are
compared. Lensing does not significantly affect the galaxy number
counts at low luminosity, while at the bright end, lensing
dramatically enhances the number counts. But all the predictions with
our model (sources at redshift $z_s=3.0$, $\sigma_8=0.8$ and $M_{\rm
  cool }=10^{13}\msun$) have lower number counts than that observed by
HerMES. \citet{2012arXiv1205.3778W} also investigated the lensing
effect on the HerMES SMG data and find that this discrepancy is mainly
due to the contamination from the late-type sprials. Although our
lensing efficiency has been enhanced by the SIS halo, the discrepancy
is still significant.
The number of lensed SMGs in which these sprial
galaxies have been subtracted \citep{2012arXiv1205.3778W} is
shown as the solid points in the figure. Our results confirms that
gravitational lensing is a natural way to explain the behaviour of SMG
luminosity function at the bright end.

In Fig.~\ref{fig:lmbig}, the shaded region shows the uncertainty due
to source redshift between $2.0$ and $4.0$. All of our predictions
with a single source redshift are larger than the lensed data (solid
points). In reality, the source will have a redshift
distribution. Different source redshift distributions are suggested
due to different selection criteria. A significant number of low
redshift sources are also found in the Herschel survey
\citep[e.g.][]{2012arXiv1210.4928C}. Furthermore, the Schechter
luminosity function parameters such as $S^*$ or $\alpha$ may be
different as well. So a more realistic prediction will depend on these
unknown parameters, nevertheless the agreement is encouraging.

In Fig.~\ref{fig:lm}, we show the effects due to cosmological
parameter $\sigma_8$. As one expects, a large $\sigma_8$ will increase
the halo number density for all masses, which will increase the lensed
galaxy number counts at both bright and faint ends. From the shaded regions
in Figs.~\ref{fig:lima}, \ref{fig:lmbig} and \ref{fig:lm}, one can see that
the uncertainty of source redshift affects our result the most.

In addition, we perform our calculation using different $\mu_{\rm
  min}$ ($1.05$ and $1.15$). The effects on the number counts are
small, $5\%$ at the bright end and $7\%$ at the faint end. In
\citet{2011A&A...536A..85H}, a log-normal approximation of lensing
probability is found using large volume numerical simulations. We use
the fitting formula given by \citet{2011A&A...536A..85H} to calculate
the probability for $\mu<1.1$. We perform the integral from $\mu_{\rm
  min}=0.01$ to $\mu_{\rm max}=30$ and find that the change to the
number counts at the bright end is also small (blue line in
Fig.~\ref{fig:lm}). The upper limits $\mu_{\rm max}$ is uncertain as
well. \citet{2002MNRAS.329..445P} consider the SMG population and
estimate $\mu_{\rm max} \sim 10-30$. However, a magnification of
$\mu\sim45$ is reported in the cluster Abell 2218
\citep{2004MNRAS.349.1211K}. Our results show that it causes a
decrease of $50\%$ at the bright end ($S>0.1$ Jy) if we adopt
$\mu_{\rm max}=20$ but keep the faint end unchanged. This effects will
be blurred if the finite source size of background galaxies is taken
into account and therefore needs a more detailed study.
%

\section{Summary and Discussion}

In this paper we have studied the lensing effects of galaxy and
cluster haloes, and how lensing will change the luminosity function of
high redshift sub-millimetre galaxies. The lensing properties for
individual haloes and haloes as a population have been presented. In
particular, we find that the halo ellipticity does not affect the
lensing efficiency significantly with our definition of ellipse
coordinate.  On the other hand, as expected, halo mass profiles
significantly affect the lensing cross section. Not surprisingly, the
NFW profile has lower lensing efficiency than the SIS profile. We
argued that a combination of two population of halo profiles is more
realistic: we use the SIS model for mass less than $M_{\rm cool}$ and
NFW for mass greater than $M_{\rm cool}$. The SIS profile is favoured
for galactic sized haloes, due to the baryonic cooling effect. It is
also close to the composite profile of an NFW dark halo and a Sersic
stellar component \citet{2007ApJ...667..176G,2012ApJ...755...46L}. Our
prediction with simple assumptions (two population, circular symmetric
halo profile) can match the observation given in
\citet{2010ApJ...717L..31L} quite well.
Compared with previous works, our model is simpler and more realistic.
For the SMG number counts of \citep{2012arXiv1203.2562H}, the excess
on the bight end can not be explained by lensing alone. The
contamination from local late-type sprial galaxies is
non-negligible. On the other hand, the cleaned lensed SMGs sample can
be easily fit by lensing. We vary $M_{\rm cool}$ between $10^{12}$
and $10^{13} \msun$, and find it only slightly affects the lensing
probability. Moreover, our results also show the dependence of the
lensing probability on the cosmological parameter, $\sigma_8$. This is
especially important for large magnifications, since in the halo
model, the number density of massive haloes is sensitive to
$\sigma_8$.

In a simple test of total cross section as a function of halo mass, we
find that most contribution to the lensing cross section comes from
dark matter haloes with mass between $10^{12}$ and $10^{15} \msun$,
i.e. massive galaxies or galaxy groups. The upper mass limit is due to the
rare number of massive haloes. It also explains why most lenses
are at low redshift, since massive haloes have not yet formed
at earlier times. The lower limit is due to the small cross sections
of individual low-mass lens haloes, because they generate little cross
sections.

Our model is simplistic, a number of improvements can be made. (1) We
did not take into account the additional matter distribution between
the source and observer, i.e. several small haloes may contribute to a
modest lensing magnification. The contribution to the cross section
appears to be small, e.g. a few percent
\citep{2007MNRAS.382..121H}. (2) We assumed the ellipticity is
constant as a function of radius. For relaxed haloes, this may be a
fair approximation \citep{2002ApJ...574..538J}. However, baryons may
make the central parts more spherical \citep{2004IAUS..220..421S}.
Even more significantly, for merging galaxies or clusters, the cross
section may be enhanced.  (3) We ignored the finite source size, which
may reduce the magnification effect \citep{2011ApJ...734...52H}. Using
a smaller maximum magnification \citep{2002MNRAS.329..445P}, the
predicted number count becomes smaller at the bright end. However, the
overall effect to the lensing probability due to finite source size
needs a more detailed study. The limited resolution of Herschel may
also confuse multiple images as a single one, and thus change the
magnification bias.  (4) The multiple images due to substructures and
external shear may increase the lensing efficiency. A more detailed
study with numerical simulations is desirable to address some of these
issues.

\section*{Acknowledgments}
We thank Julie Wardlow for providing us the HerMES data, Jin An, Awat
Rahimi, Chuck Keeton, Lin Yan, Hai Fu and Marcos Lima for useful comments on the
draft, Yixian Cao and Liang Gao for computing support. XE is supported
by NSFC grant No.11203029. GL is supported by the One-Hundred-Talent
fellowships of CAS and by the NSFC grant (No.11243005 and
11273061). SM is supported by the Chinese Academy of Sciences and
National Astronomical Observatories. LC is supported by the Young
Researcher Grant of National Astronomical Observatories and NSFC
grant No.11203028.

\bibliographystyle{mn2e}
\bibliography{../../../bib/refbooks,../../../bib/lens,../../../bib/bhlens,../../../bib/refcos,../../../bib/smg,../../../bib/galaxy,../../../bib/stronglens,../../../bib/shape,../../../bib/flexion,../../../bib/before70}

\appendix
\section{Cross section of SIE and NIE haloes}
From \citet{1998ApJ...495..157K},  the surface density for the NIE profile is
\be
\kappa(x,y) = {b_{\rm I} \over 2 \sqrt{q^2(x^2+s^2) + y^2}}.
\elabel{ekappa0}
\ee
Its magnification reads
\be
\mu^{-1}=1-{b_{\rm I} \over \psi} +{{b_{\rm I}}^2s \over \psi[(\psi+s)^2+(1-q^2)x^2] },
\ee
where $\psi^2=q^2(s^2+x^2+y^2)$.  Now we define an elliptical
coordinate, $\Theta=\sqrt{q\theta_1^2+\theta_2^2/q}$, and write the
surface density as
\be
\kappa(\Theta) = {\theta_{\rm E} \over 2 \sqrt{\Theta^2 + \theta_c^2}}.
\elabel{ekappa}
\ee
These definitions keep the total mass within an ellipse invariant with ellipticity, $q$ for given $\Theta_{\rm E}, \theta_c$ and $\Theta$. We can rewrite Eq.~\ref{eq:ekappa} as
\be
\kappa = {q\theta_{\rm E} \over 2 \sqrt{q^2[(\sqrt{q}\theta_1)^2+\theta_c^2] + (\sqrt{q}\theta_2)^2 }}.
\elabel{ekappat}
\ee
Comparing Eqs.~\ref{eq:ekappa0} and \ref{eq:ekappat}, the magnification in our definition is
\be
\mu^{-1}(\Theta,x)=1-{\theta_{\rm E} \over \Psi} +{{\theta_{\rm E}}^2\theta_c \over \Psi[q(\Psi+\theta_c/q)^2+(1-q^2)\theta_1^2] },
\elabel{emag}
\ee
where $\Psi=\sqrt{\Theta^2+\theta_c^2}$. The cross section for a given
magnification threshold is defined as
\be
\sigma(\mu_{\rm min}) =\int\int_{|\mu| >\mu_{\rm min}} \d^2 \beta =
\int \int_{|\mu| >\mu_{\rm min}} {1 \over |\mu|} \;\d^2 \theta
\elabel{ecross}
\ee
and it can always be calculated by numerical integration.
The SIE model has $\theta_c=0$ and has
 an ellipsoidal magnification distribution. The corresponding cross section
can be derived analytically (see Eqs.~\ref{eq:siemag} and \ref{eq:crosssie}).

\section{Numerical Method for elliptical halo lensing properties}
The lensing properties of an elliptical halo can be calculated numerically
given an arbitrary surface density profile \citep{2001astro.ph..2341K}.
Scaling coordinates as $x=\sqrt{q}\theta_1$ and $y=\sqrt{q}\theta_2$, we
can rewrite the ellipsoidal distribution as
\be
\kappa=\kappa(\Theta), \;\;\; {\rm where}\;\;\;
\Theta^2=q\theta_1^2+\theta_2^2/q = x^2 + y^2/q^2.
\ee
The lensing properties for a surface density distribution with
elliptical symmetry can be written as a set of one-dimensional
integrals,
\bea
\psi(x,y)     &=& {q\over 2} I(x,y)\\
\psi_x(x,y)   &=& qxJ_0(x,y)\\
\psi_y(x,y)   &=& qyJ_1(x,y)\\
\psi_{xx}(x,y) &=&  q x^2 K_0(x,y) + qJ_0(x,y) \\
\psi_{yy}(x,y) &=&  q y^2 K_2(x,y) + qJ_1(x,y) \\
\psi_{xy}(x,y) &=&  q xy K_1(x,y).
\eea
Here
\bea
I(x,y)  &=&  \int_0^1 \frac{\xi(u)\psi_r (\xi (u)) }{u\eck{1-(1-q^2)u}^{1/2}}\d u,\\
J_n(x,y) &=& \int_0^1 \frac{\kappa (\xi (u)) }{\eck{1-(1-q^2)u}^{n+1/2}}\d u,\\
K_n (x, y) &=& \int_0^1
\frac{u\,\kappa'\,(\xi (u) )\,} {\xi(u)[1 - (1 - q^2 )u]^{n+1/2}}\d u,
\eea
are one-dimensional integrals, where $\psi_r$ is the circular deflection angle,
 $q$ is the axis ratio of the lens
and $\kappa'$ are the first order derivatives of the convergence,
i.e. $\kappa'(\xi) = \d \kappa(\xi) / \d \xi$.  The convergence
$\kappa$ is written as a function of the ellipse coordinate $\xi(u)$
given by
\be
\xi(u)^2 = u \rund{x^2 + {y^2 \over 1- (1-q^2) u}}.
\ee
Therefore once we know the behavior of $\kappa'(r)$ and $\psi_r(r)$ of
a circular symmetric $\kappa(r)$, the lensing properties of the
elliptical mass distribution $\kappa(\Theta)$ can always be obtained
through these integrals. The above equations are slightly different
from those in \citet{2001astro.ph..2341K}. Considering the scaling
relation between coordinates ($x, y$) and ($\theta_1$, $\theta_2$), we
simply have
\bea
\psi(\theta_1,\theta_2)     &=& \psi(x,y)/q\\
\psi_{\theta_1}(\theta_1,\theta_2)   &=& \psi_x(x,y)/\sqrt{q}\\
\psi_{\theta_2}(\theta_1,\theta_2)   &=& \psi_y(x,y)/\sqrt{q}\\
\psi_{\theta_1 \theta_1}(\theta_1,\theta_2) &=&  \psi_{xx}(x,y)\\
\psi_{\theta_1 \theta_2}(\theta_1,\theta_2) &=&  \psi_{xy}(x,y)\\
\psi_{\theta_2 \theta_2}(\theta_1,\theta_2) &=&  \psi_{yy}(x,y).
\eea

\end{document}